\documentclass[
 superscriptaddress,
 amsmath,amssymb,
 aps,
prb,
floatfix,
twocolumn,
10pt,longbibliography
]{revtex4-2}

\usepackage{graphicx}
\usepackage{braket}

\usepackage{xcolor}

\usepackage{orcidlink}

\newcommand{\nep}{e}
\newcommand{\ud}{d}
\newcommand{\bruket}[1]{\overline{(#1)}}

\newcommand{\opcdag}[1]{{\hat{c}^{\dagger}}_{#1}}

\newcommand{\opc}[1]{{\hat{c}^{\phantom \dagger}}_{#1}}

\hypersetup{%
	colorlinks=true, 
	breaklinks=true,
	urlcolor=blue, 
	linkcolor=blue, 
	citecolor=blue,
}

\begin{document}

\title{Thermalization propagation front and robustness against avalanches in localized systems}

\author{Annarita Scocco\,\orcidlink{0000-0002-6920-5843}}
\affiliation{Scuola Superiore Meridionale, Largo San Marcellino 10, I-80138 Napoli, Italy}
 
\author{Gianluca Passarelli\,\orcidlink{0000-0002-3292-0034}}
\affiliation{Dipartimento di Fisica ``E. Pancini'', Universit\`a di Napoli Federico II, Complesso di Monte S. Angelo, via Cinthia, I-80126 Napoli, Italy}

\author{Mario Collura\,\orcidlink{0000-0003-2615-8140}}
\affiliation{SISSA, Via Bonomea 265, I-34136 Trieste, Italy}
\affiliation{INFN, Sezione di Trieste, Via Valerio 2, 34127 Trieste, Italy}

\author{Procolo Lucignano\,\orcidlink{0000-0003-2784-8485}}
\affiliation{Dipartimento di Fisica ``E. Pancini'', Universit\`a di Napoli Federico II, Complesso di Monte S. Angelo, via Cinthia, I-80126 Napoli, Italy}

\author{Angelo Russomanno\,\orcidlink{0009-0000-1923-370X}}
\affiliation{Dipartimento di Fisica ``E. Pancini'', Universit\`a di Napoli Federico II, Complesso di Monte S. Angelo, via Cinthia, I-80126 Napoli, Italy}

\begin{abstract}

We investigate the robustness of the many-body localized (MBL) phase to the quantum-avalanche instability by studying the dynamics of a localized spin chain coupled to a $T=\infty$ thermal bath through its leftmost site. By analyzing local magnetizations we estimate the size of the thermalized sector of the chain and find that it increases logarithmically slowly in time. This logarithmically slow propagation of the thermalization front allows us to lower bound the slowest thermalization time, and find a broad parameter range where it scales fast enough with the system size that MBL is robust against thermalization induced by avalanches. The further finding that the imbalance -- a global quantity measuring localization -- thermalizes over an exponential time scale both in disorder strength and system size is in agreement with these results.

\end{abstract}

\maketitle

\section{\label{sec:intro} Introduction}

Thermalization in quantum systems occurs in a way remarkably different than in classical systems, by a mechanism called eigenstate thermalization hypothesis (ETH)~\cite{deutsch1991quantum,PhysRevE.50.888,PhysRevE.60.3949}. In a thermalizing quantum system eigenstates are locally equivalent to thermal density matrices, and this gives rise to long-time thermalization of local observables~\cite{polkovnikov2011colloquium,d2016quantum,rigol2008thermalization}. Generic isolated quantum systems are expected to thermalize and obey ETH~\cite{Reimann_2016}. It is therefore of particular interest to find systems that avoid thermalization: In this case, quantum information encoded in the initial state can persist for long times, with relevance for technological applications, as quantum memories \cite{serbyn2014interferometric}. 

In an ergodic system, thermalization occurs because its various parts can exchange particles and energy, so a possible way for a system to avoid thermalization is to exhibit insulating behavior, an example of which is given by Anderson localization~\cite{anderson1958absence}, occurring in non-interacting disordered systems. In the presence of small enough interactions, the system is still space localized and non thermalizing, a phenomenon called many-body localization (MBL)~\cite{fleishman1980interactions}. Following the works~\cite{basko2006metal,gorni2005interacting}, this topic has been tremendously explored over the past few years, from a theoretical~\cite{ronen2013,vosk2015theory,imbrie2016on}, experimental~\cite{schreiber2015observation,choi2016exploring,smith2016programmable,bordia2017periodically,lukin2019probing,rubio2019many}, and numerical~\cite{oganesyan2007localization,vznidarivc2008many,pal2010many,bauer2013area,DeLuca_2013,kjall2014many,luitz2015many,PhysRevB.105.184202,PRXQuantum.5.020352,lev2015absence,iemini2016signatures,doggen2021168437} point of view, particularly focusing on one-dimensional systems.
MBL systems display many interesting features such as the emergence of a complete set of quasi-localized integrals of motion~\cite{serbyn2013local,huse2014phenomenology,imbrie1017local}, area law entanglement in all many-body eigenstates~\cite{de2006entanglement,eisert2010colloquium,bauer2013area}, logarithmic growth of entanglement entropy with time~\cite{Moore_PRL12,Serbyn_PRL13,vznidarivc2008many,de2006entanglement}. This last slow-entanglement-growth property is especially relevant, due the role played by entanglement in giving rise to ETH of local observables~\cite{Roy_2015}. The properties of MBL systems have been extensively described in different reviews~\cite{alet2018many,abanin2019colloquium,sierant2024many}. 

However, the stability of this {regime} has been put into question in the thermodynamic limit, as it was pointed out that under certain circumstances many-body localized systems may be unstable towards rare regions of small disorder by a mechanism dubbed ``quantum avalanches''~\cite{de2017stability}. {This phenomenon has been considered in} one- or higher-dimensional systems, both theoretically~\cite{PhysRevB.99.134305,agarwal2017rare,potirniche2019exploration,szoldra2024catching} and experimentally~\cite{leonard2023probing,luschen2017signatures}, studying the spectral properties or the dynamics of a MBL system in contact with an ergodic inclusion~\cite{luitz2017small,colmenarez2024ergodic,peacock2023many}. 
In some important cases the effect of the ergodic inclusion was studied using a Lindbladian acting on an end of the system~\cite{PhysRevB.106.L020202,Tu_2023,PhysRevB.105.174205}, as we better clarify below.~\cite{nota_the}  

The search for evidence of the avalanche mechanism in standard MBL models is still very active.
One possible approach is through many-body resonances, which allow globally different spin configurations to interact~\cite{PhysRevB.92.104202,PhysRevB.105.174205,Crowley_SciPostPhys22}. They are negligible in the MBL phase but there is a crossover regime where they become more and more relevant with increasing system size. This leads eventually to avalanches~\cite{PhysRevB.105.174205}, that spread just thanks to {the coupling} between rare near-resonant eigenstates~\cite{PhysRevLett.130.250405}. {Avalanches are also} related to interaction-driven instabilities seen in the behavior of correlation lengths~\cite{colbois2024}. From the experimental point of view, in Ref.~\cite{leonard2023probing}, {avalanche spread is monitored} by measuring the site-resolved entropy over time.

{The studies mentioned above argue in favor of the robustness of MBL to avalanches, although robustness occurs above a disorder-strength threshold larger than the one corresponding to the finite-size crossover to MBL. There are also critical voices to the existence of MBL in the thermodynamic limit. They report a crossover point to MBL that linearly increases with the system size. In~\cite{PhysRevE.102.062144,PhysRevB.102.064207} this result is obtained by numerically analyzing the ratio of the Thouless time and the Heisenberg time obtained from the spectral form factor, and in~\cite{PhysRevE.104.054105} by numerically studying the behavior of the fidelity susceptibility. These are all numerical finite-size results, so the question is still debated, and far from being settled, because finite-size numerical results cannot be univocally extrapolated to the thermodynamic limit~\cite{sierant2024many}. }

{Our contribution comes into this debate by looking from a different perspective at}
the idea that an ergodic inclusion, which occurs almost certainly in sufficiently large systems, can thermalize the entire chain if the thermalization time of any subchain is short enough~\cite{PhysRevB.105.174205}. This time can be numerically estimated by coupling a subchain to a thermal bath by one of its ends, simulating the already thermalized part of the chain. If the slowest thermalization time, corresponding to the time it takes for the farthest spin to thermalize, scales slowly enough with the size of the subchain, then the avalanche propagates and the system thermalizes~\cite{PhysRevB.105.174205,PhysRevB.106.L020202}.

In this work, {we suggest a different way to estimate this thermalization time, based on the logarithmically slow propagation of the heat.} We couple an MBL system to a $T=\infty$ thermal bath by one of its extremities and study how the slowest thermalization time scales with the system size, to understand if MBL is robust to avalanches. We use the {single-site} Lindbladian bath considered in~\cite{Talia_Bar_Lev_22}, applying it to the leftmost site, as in~\cite{PhysRevB.106.L020202,Tu_2023,PhysRevB.105.174205}
This choice of bath can be numerically studied with an approach of Hamiltonian dynamics with noise~\cite{Talia_Bar_Lev_22,devendra_24,DeLuca2019}, allows scaling to large system sizes in the case of Anderson localized system, and can provide a lower bound to the slowest thermalization time. Furthermore, it allows scaling to large system sizes in the case of Anderson localized systems. 

\begin{figure}[t]
\centering
\includegraphics[width=\columnwidth]{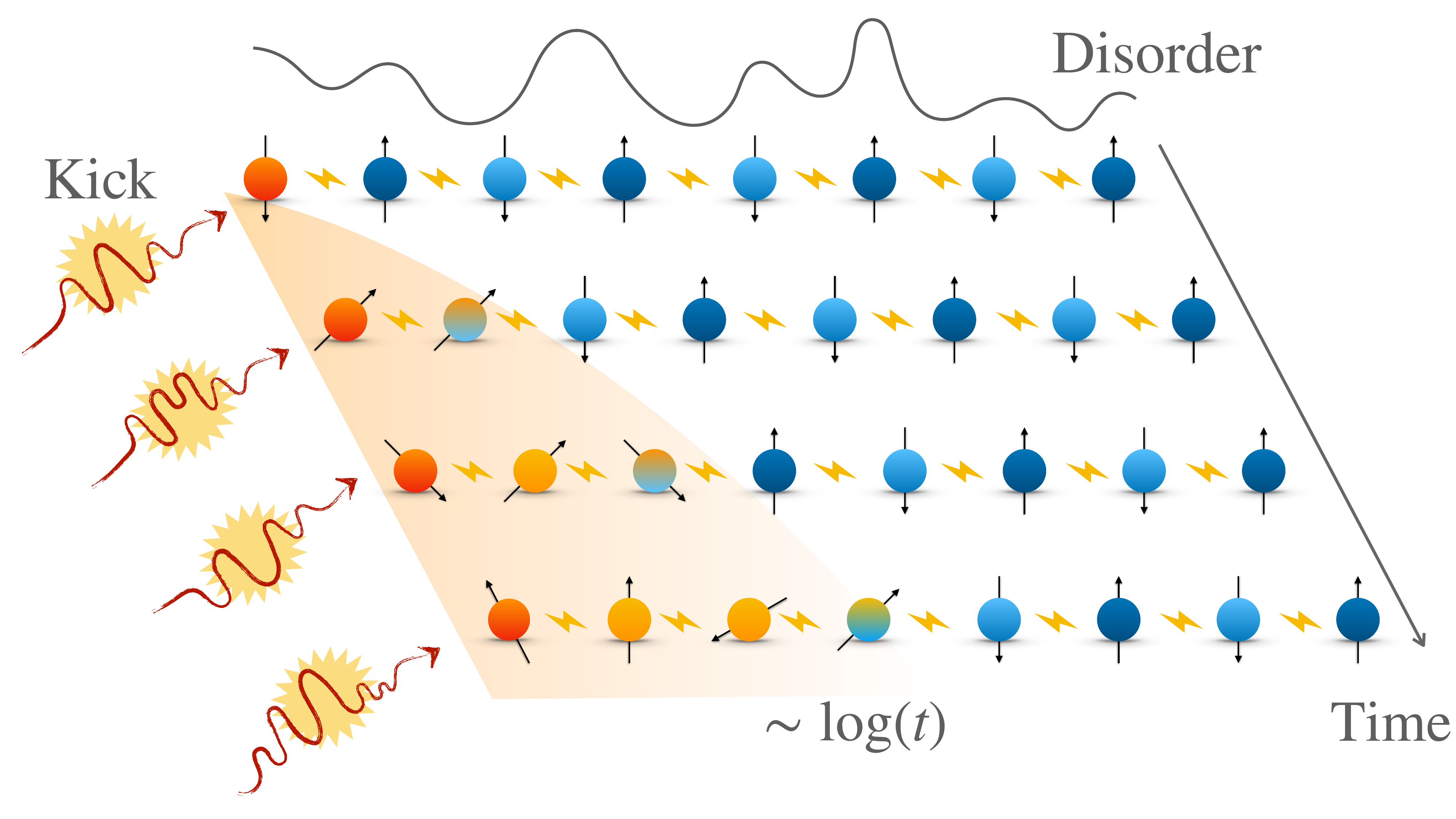}
\caption{Graphical visualization of the model: a 1D chain of interacting 1/2 spins with a fixed disorder profile, starting from the Néel state, and with open boundary conditions. Different copies of the system mean different subsequent times, as indicated by the time arrow on the right. The leftmost spin is coupled to a $T=\infty$ thermal bath (a noisy time-dependent kick) and a thermalization front (shaded area) propagates logarithmically slowly.}
\label{fig:sketch}
\end{figure}

We {estimate} the slowest thermalization time by looking at the propagation of the thermalization front through the chain, an approach that, to the best of our knowledge, has not yet been used in this context. Heat propagates through the system, from the bath at the leftmost site, and at any time there is a subchain on the left side that has already thermalized [see the cartoon in Fig.~\ref{fig:sketch}]. We estimate the length of this thermalized subchain by defining a thermalization length scale based on the behavior of the local magnetizations. 
We find that both in Anderson and MBL localized systems this length scale increases logarithmically in time {-- in agreement with the general analytical predictions of logarithmic light cones in MBL systems of~\cite{kim2014localintegralsmotionlogarithmic,elgart2024slowpropagationinformationrandom}, the analytical prediction of their robustness under local perturbations of~\cite{toniolo2024}, and the logarithmic increase of a length scale related to the two-time density-density correlators in the Anderson case~\cite{Talia_Bar_Lev_22}} undergoing single-site noise.

We use the logarithmic growth of the thermalization length scale to estimate the time in which the thermalization front reaches the other end of the chain. We find that this time scale exponentially increases with the system size, and there is a regime of parameters where this exponential increase is fast enough that localization is robust against thermalization induced by avalanches. {We find that the slope of the logarithmic increase of the thermalization length scale is the same both with a particle-number-conserving or a particle-number nonconserving coupling to the bath, consistently with the logarithmic light cone being independent of this conservation~\cite{toniolo2024}. We remark also that the results we find for the Anderson-localized model provide information relevant also for MBL, as the coupling to the bath makes the system interacting~\cite{luitz2017small}.} We also study the slowest thermalization time by looking at the thermalization behavior of the imbalance and obtain results in agreement.

The paper is organized as follows. In Sec.~\ref{model:sec} we define our model, and in Sec.~\ref{sec:methods} we discuss the numerical methods we use in the case of many-body localization. In Sec.~\ref{imba_rob:sec} we numerically show that the imbalance thermalization time is exponential both in the system size and in the disorder strength, and this allows us to show that there is a critical strength beyond which MBL is robust against the avalanche instability. In Sec.~\ref{therma_front:sec} we define a thermalization length scale using the local magnetizations and show that it logarithmically increases with time. This implies that the thermalization time of the imbalance exponentially increases with system size. 
In Sec.~\ref{conc:sec} we draw our conclusions and sketch perspectives of future work. In the Appendixes we discuss some important aspects that would have broken the main discussion. In particular, in Appendix~\ref{derivation:sec} we review how the Hamiltonian quantum-trajectory approach works, in Appendix~\ref{anderson:sec} we discuss how the Gaussian nature of the state in the case of the Anderson model simplifies the numerical analysis, {in Appendix~\ref{relax:app} we take a particle-number nonconserving bath and find that the logarithmic propagation occurs with the same slope}, and in Appendix~\ref{ava_come_lava:sec} we review the derivation of the threshold above which the scaling of the slowest thermalization time is slow enough to guarantee robustness against avalanches. 

\section{\label{sec:mode} Model}\label{model:sec}

We consider the 1D spin-1/2 XXZ Heisenberg model in a random magnetic field 
\begin{equation}
     \hat{H}_0 = \sum_{j=1}^L h_j \hat{S}_j^z +  J \sum_{j=1}^{L-1} ( \hat{S}^x_j \hat{S}^x_{j+1} + \hat{S}^y_j \hat{S}^y_{j+1} + \Delta \hat{S}^z_j \hat{S}^z_{j+1} ),
     \label{eq:xxzmodel}
\end{equation}
where $L$ is the system size, $\hat{S}^{\alpha}_j \equiv \hat{\sigma}_j^{\alpha}/2$ are the on-site magnetizations ($\alpha \in \lbrace x, y, z \rbrace$) and $J = 1$ sets the energy scale. The on-site magnetic fields $h_j$ are chosen randomly and uniformly in the interval $[-W, W]$, and the system conserves the total magnetization $S^z_{\rm tot}=\sum_{j=1}^L\hat{S}^z_j$ in the $z$ direction. We consider open boundary conditions. 

This model has been widely studied as a paradigmatic model displaying MBL behavior~\cite{luitz2015many,sierant2022challenges,vznidarivc2008many,colmenarez2019statistics,sierant2019level,chanda2020time,PhysRevE.102.062144,serbyn2016power,sierant2022challenges}. It exhibits {at finite sizes a crossover}~\cite{PhysRevB.105.174205} from the ergodic to the localized regime, and it is believed to capture the essential properties of the MBL systems. 
At $\Delta = 1$ various estimates of {this crossover point $W_c$ have been made, some of them including $W_c \approx 3.5$~\cite{pal2010many}, $W_c \approx 3.7$~\cite{luitz2015many}, $W_c \approx 5$~\cite{doggen2018many}, $W_c \approx 5.4$~\cite{sierant2020polynomially}, nevertheless it is impossible to univocally extrapolate finite-size results to the thermodynamic limit, and so the question is far from being settled~\cite{oganesyan2007localization,sierant2024many,PhysRevE.102.062144,PhysRevB.102.064207,PhysRevE.104.054105,ABANIN2021168415}.}
Furthermore, the Jordan-Wigner transformation~\cite{lieb_AP61} allows to map this model to a system of interacting spinless fermions with tunneling matrix element $J$ and nearest-neighbor interaction strength $\Delta$ (see Appendix~\ref{anderson:sec} for details), connecting the model and the experiments on quasirandom optical lattices~\cite{schreiber2015observation}. 

We couple one end of this system to a thermal bath that induces $T=\infty$ thermalization. This evolution is described by the Lindbladian 
\begin{equation}\label{lindblad:eqn}
    \dot{\hat{\rho}}(t) = - i [ \hat{H}_0, \hat{\rho}(t) ] - \gamma^2 [\hat{S}^z_1 ,[\hat{S}^z_1, \hat{\rho}(t)]]\,,
\end{equation}
where $\gamma^2$ is the coupling to the bath. {This noise conserves the value of $S^z_{\rm tot}$, corresponding to the particle number in the fermionic representation. (We relax this particle-number-conserving property in Appendix~\ref{relax:app}, finding essentially no difference for the slope of the logarithmic propagation discussed in Sec.~\ref{therma_front:sec}.}) Restricting to any $S^z_{\rm tot}$ subspace, one can see that the identity is the only steady state of this Lindblad dynamics, implying the $T=\infty$ thermalization inside that subspace. 

In order to study this Lindbladian, we use a quantum trajectory approach. More specifically we rely on the so-called unitary unraveling~\cite{DeLuca2019,Talia_Bar_Lev_22} where the dynamics of Eq.~\eqref{lindblad:eqn} is described by an average over many realizations of unitary Schr\"odinger evolutions with a noisy Hamiltonian. So one should evolve with the stochastic Hamiltonian
\begin{equation}\label{hmbl:eqn}
  \hat{H}_{\gamma}(t) = \hat{H}_0 + \gamma\xi(t)\hat{S}_1^z\,,
\end{equation}
where $\hat{H}_0$ is defined in Eq.~\eqref{eq:xxzmodel} and $\xi(t)$ is an uncorrelated Gaussian noise, for which $\braket{\xi(t)} = 0$, $\braket{\xi(t)\xi(t')}=\delta(t-t')$ and all the cumulants are vanishing. Angular brakets denote the average over noise realizations. {This Hamiltonian for the Anderson $\Delta=0$ case has already been considered in~\cite{Talia_Bar_Lev_22}, where the entanglement entropy and the two-time density-density correlations are considered.} In order to numerically  implement this evolution, we must discretize it over time intervals $\tau\ll 1/J$ and Trotterize it, so that the $n$-th evolution step is given by the action of the operator
\begin{equation}\label{unno:eqn}
  \hat{U}_n = \nep^{-i\eta_n\gamma\hat{S}_1^z}\nep^{-i\tau\hat{H}_0}\,,
\end{equation}
where $\eta_n$ are Gaussian random variables defined as $\eta_n = \int_{(n-1)\tau}^{n\tau} \xi(t) dt$, so that $\braket{\eta_n}=0$ and $\braket{\eta_n\eta_{n'}} = \tau \delta_{n,n'}$. In the following we fix $\tau = 0.05$, as we have verified that a shorter $\tau$ does not affect the result. Averaging over random realizations, in the limit $\tau\to 0$ one recovers the Lindblad equation, Eq.~\eqref{lindblad:eqn}, as we show in Appendix~\ref{derivation:sec}. In the Anderson-model case $\Delta = 0$ the dynamics along each trajectory is given by Gaussian states, allowing thereby to numerically address large system sizes {, as we discuss in detail in Appendix~\ref{anderson:sec}}.

\section{Methods}\label{sec:methods}

For the dynamics, it is customary to evolve the system starting from the high energy, out-of-equilibrium, unentangled, antiferromagnetic Néel state $\ket{\psi} = \ket{ \uparrow \downarrow \dots \uparrow \downarrow} $, which has total magnetization $S_{\rm tot}^z = 0$ and gives rise to a dynamics restricted to the corresponding $S^z_{\rm tot}$ subspace whose dimension is $\mathcal{N} = \binom{L}{L/2}$. 
In order to detect the effect of the noise on the system, we compare two different evolutions, one with the Hamiltonian Eq.~\eqref{eq:xxzmodel}, without noise ($\gamma=0$)
\begin{equation}
    \ket{\psi_0(t)} = e^{- i \hat{H}_0 t} \ket{\uparrow \downarrow \uparrow \downarrow ...}\,,
    \label{eq:stategamma0}
\end{equation}
and one with the fully noisy Hamiltonian $\hat{H}_{\gamma}$ [Eq.~\eqref{hmbl:eqn}], fixing the same disorder realization
\begin{equation}
    \ket{\psi_{\gamma}(t)} = \prod_{n=1}^{[t/\tau]} \hat{U}_n \ket{\uparrow \downarrow \uparrow \downarrow ...}\,,
    \label{eq:stategamma}
\end{equation}
ending both evolutions at some final time $t_f$. The state $\ket{\psi_0(t)}$ is unaffected by noise and will be considered as a reference. We average each of the quantities over $N_{r}$ different disorder/noise realizations. When $\gamma=0$ in each realization we take a different random choice of the onsite fields $h_j$, for $j\in\{1,\,\ldots,\,L\}$ in Eq.~\eqref{eq:xxzmodel}. When $\gamma=1$ we take in each realization a different choice of the $h_j$ and also a different choice of the random sequence $\eta_n$, with $n\in\{1,\ldots,[t_f/\tau]\}$ providing the noise in Eq.~\eqref{unno:eqn}. We indicate the average over the $N_r$ disorder/noise realizations with an overline $\overline{(\ldots)}$. We evaluate the errorbar on this average as the root mean square deviation divided by $\sqrt{N_{r}}$, performing error propagation where appropriate.

In the case of the Anderson model we take $N_r = 1200$ while in the case of MBL, in which the simulation times are longer, the number of realizations will be going from $N_r = 500$ for the smallest system $L = 8$, to $N_r = 100$ for the biggest one $L = 16$. When we compare $\gamma=1$ and $\gamma = 0$ {we average over the same set of disorder realization}.  To simulate the time evolution we use exact diagonalization and Krylov subspace projection methods~\cite{EXPOKIT}. {Exact diagonalization is much more efficient in the case of the Anderson model ($\Delta=0$) thanks to the mapping of each trajectory to a free fermion model (see Appendix~\ref{anderson:sec}). In  the following text, whenever we consider an interacting case, we fix $\Delta=1$, and when we consider coupling to the thermal bath we take $\gamma=1$.}

\section{Imbalance and local magnetizations}\label{sec:imbalance}
%
\subsection{Imbalance behavior and robustness against avalanche instability}\label{imba_rob:sec}
The imbalance $\mathcal{I}$ between the even and odd sites in the spin representation is defined as 
\begin{equation}
    \mathcal{I}(t) = \frac{1}{L} \sum_{j=1}^L (-1)^j \langle \psi(t)|\hat S_j^z|\psi(t) \rangle.
\end{equation}
It is a global feature of the system which can be computed from local quantities, and can be experimentally measured~\cite{schreiber2015observation,bordia2016coupling}. 
The normalization ensures $\mathcal{I}(t=0) = 1$. 
The long-time stationary value of the imbalance effectively serves as an order parameter of the MBL phase, which is why it has been widely used in the literature~\cite{luitz2016extended,sierant2022challenges,doggen2018many}. {For small $W$, in the ergodic regime, a power-law decay $\mathcal{I}(t) \propto t^{-\beta}$ 
has been observed. For larger $W$, in the finite-size localized regime, one numerically sees convergence to a non-vanishing value at long times, although the problem is particularly challenging~\cite{sierant2022challenges}.}

\begin{figure}
    \centering
    \includegraphics[width=\columnwidth]{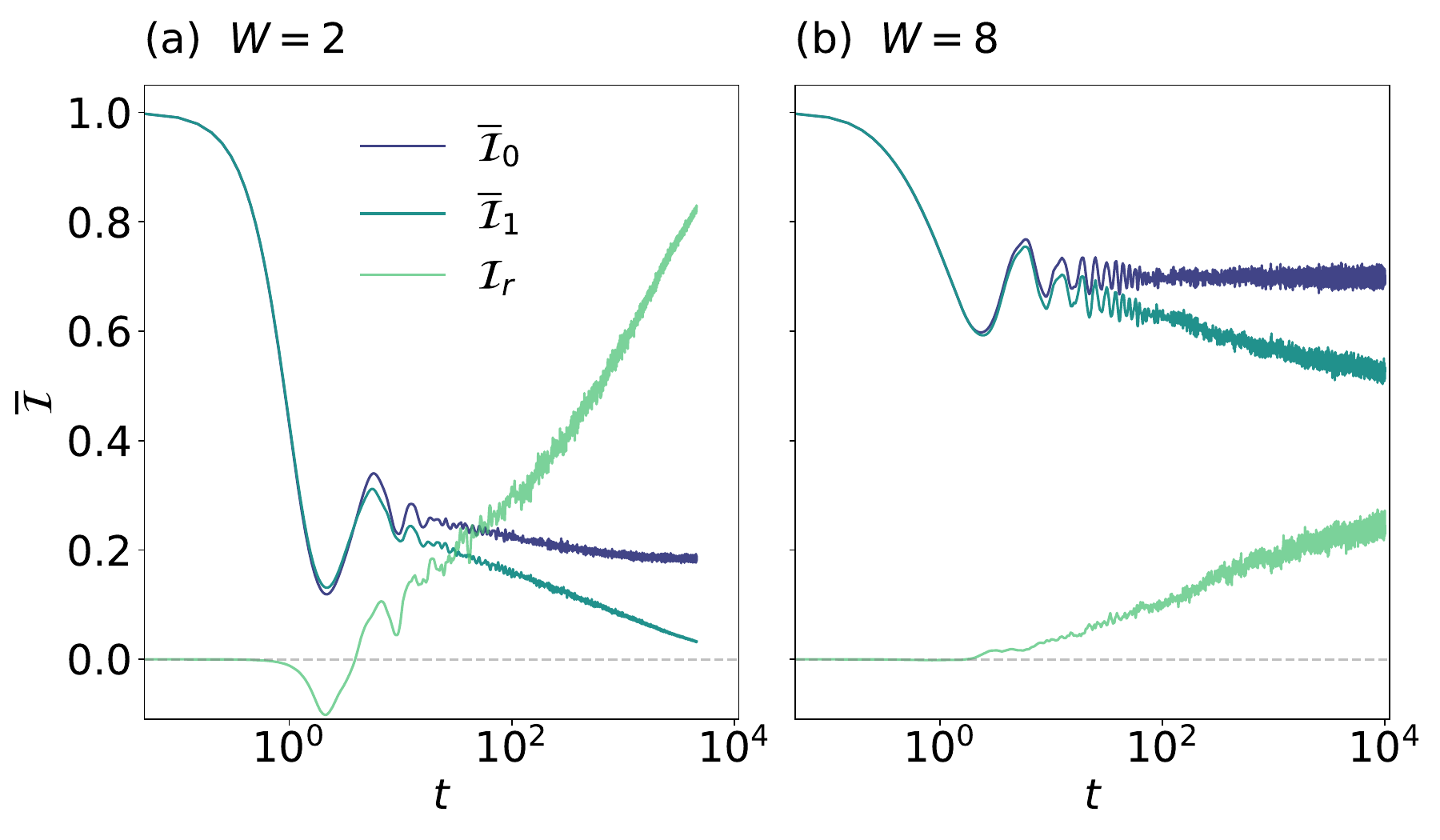}
    \caption{Interacting case. Time evolution of the imbalance in the noiseless case $\overline{\mathcal{I}}_0$, in the noisy case $\overline{\mathcal{I}}_1$ and relative difference $\mathcal{I}_r$ for $L = 16$, for $W = 2$ in panel (a) and $W = 8$ in panel (b), and $N_r = 100$.}
    \label{fig:imb_summary}
\end{figure}

The goal of our analysis is to understand how the imbalance with the noise $\overline{\mathcal I}_{\gamma = 1}(t)$ differs from the noiseless case, $\overline{\mathcal I}_{\gamma = 0}(t)$, varying the size of the system $ L $ and the strength of the disorder $W$. 
We show some examples of $\overline{\mathcal I}_{\gamma}(t)$ versus $t$ in Fig.~\ref{fig:imb_summary}, for $L=16$ and two different values of $W$ [$W=2$ in panel (a) and $W = 8$ in panel (b)].
We see that when the coupling to the bath $\gamma\neq 0$, $\overline{\mathcal{I}}_{\gamma}(t)$ deviates from $\overline{\mathcal{I}}_0(t)$,  starting at a time $t \sim 1$, and eventually decaying to zero. 

In order to perform a size scaling we define a relative imbalance $\mathcal{I}_r$ as
\begin{equation}\label{ira:eqn}
    \mathcal{I}_r(t) = \frac{\overline{I}_{0}(t) - \overline{I}_{\gamma}(t)}{\overline{I}_{0}(t) }\,.
\end{equation}
{This quantity increases from the initial value $\mathcal{I}_r(t=0)=0$ [due to $\overline{\mathcal I}_{0}(t=0) = \overline{\mathcal I}_{\gamma}(t=0)$] to the asymptotic value $\mathcal{I}_r(t\to\infty)=1$ [due to $\overline{\mathcal I}_{\gamma}(t\to\infty)=0$], as we show in Fig.~\ref{fig:imb_summary}. We can study the time scale over which this happens and we do that} computing the minimum time $\widetilde{t}$ it takes for the relative imbalance $ \mathcal{I}_{r} $ to grow above a fixed threshold, within the statistical error. Given the computational limitations, we fixed the common threshold to be $r_{\rm th}=0.17$.~\cite{nothreshold:note}

\begin{figure}
    \includegraphics[width=\columnwidth]{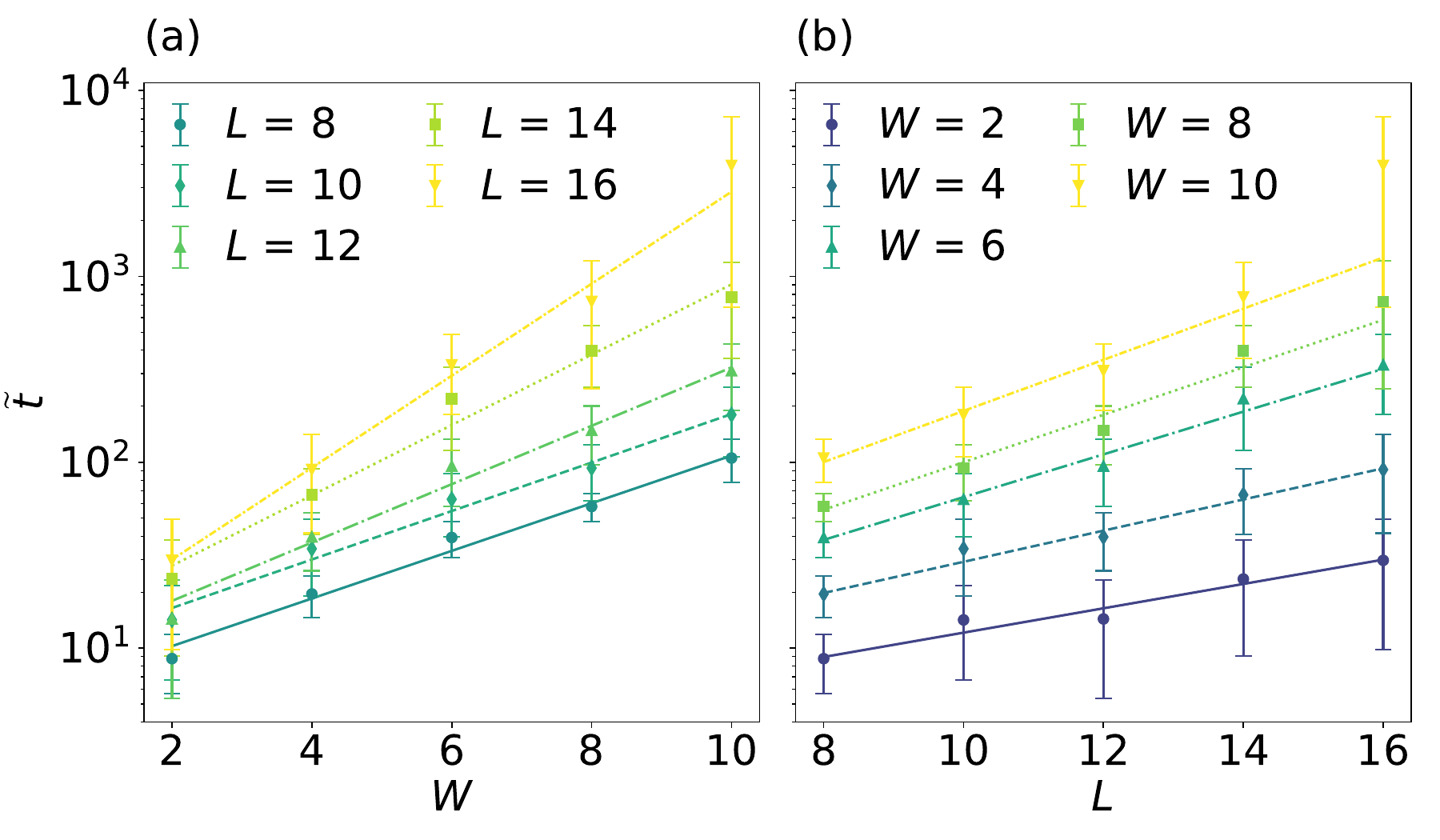}
    \caption{Interacting case. (Panel a) $\widetilde{t}$ versus $ W $ for different values of $ L $. (Panel b) $\widetilde{t}$ as a function of $ L $ for different values of $ W $. The error bars are the maximum error, {$N_{r} $} goes from $500$ for $L = 8$ to $100$ for $L = 16$.}
  \label{fig:relative}
\end{figure}

We show results in Fig.~\ref{fig:relative}. From one side we see that $\widetilde{t}$ displays a behavior consistent with an exponential increase with the disorder strength $W$, as the fits in Fig.~\ref{fig:relative}(a) show. This is in agreement with the exponential dependence of the thermalization time on the disorder strength, seen when one end of an MBL chain is time-periodically coupled to an ergodic one~\cite{peacock2023many}. 
From the other side, we find that $\widetilde{t}$ also exponentially increases with the size $L$ of the system, as we show in Fig.~\ref{fig:relative}(b).

So we conclude that, at least for the parameters considered in Fig.~\ref{fig:relative}, there is some $\alpha > 0 $ such that the thermalization time of the imbalance has the behavior
\begin{equation}\label{time:eqn}
  \widetilde{t} \sim \exp(\alpha LW)\,.
\end{equation}
We use the argument of Ref.~\cite{PhysRevB.105.174205}, according to which MBL is robust if the slowest thermalization time $t_s$ grows faster than $4^L$ (see Appendix~\ref{ava_come_lava:sec} for more details). Imposing $t^*> 4^L$ and considering that $\widetilde{t}$ is a lower bound to the slowest thermalization time $t_s$ -- as by definition $t_s > \widetilde{t}$ -- we find that there is a minimum value of disorder
\begin{equation}\label{wutan:eqn}
    \widetilde{W} = \frac{2\ln 2}{\alpha}
\end{equation}
such that for $W > \widetilde{W}$ MBL is robust against avalanche instability. We underline that MBL could still be robust even for $W < \widetilde{W}$, because $t_s$ might still grow faster than $4^L$.
If the scaling behavior observed in Fig.~\ref{fig:relative} holds over a sufficiently wide range of parameters, there exists a disorder strength beyond which MBL remains robust.

\begin{figure}
\centering
\includegraphics[width=\columnwidth]{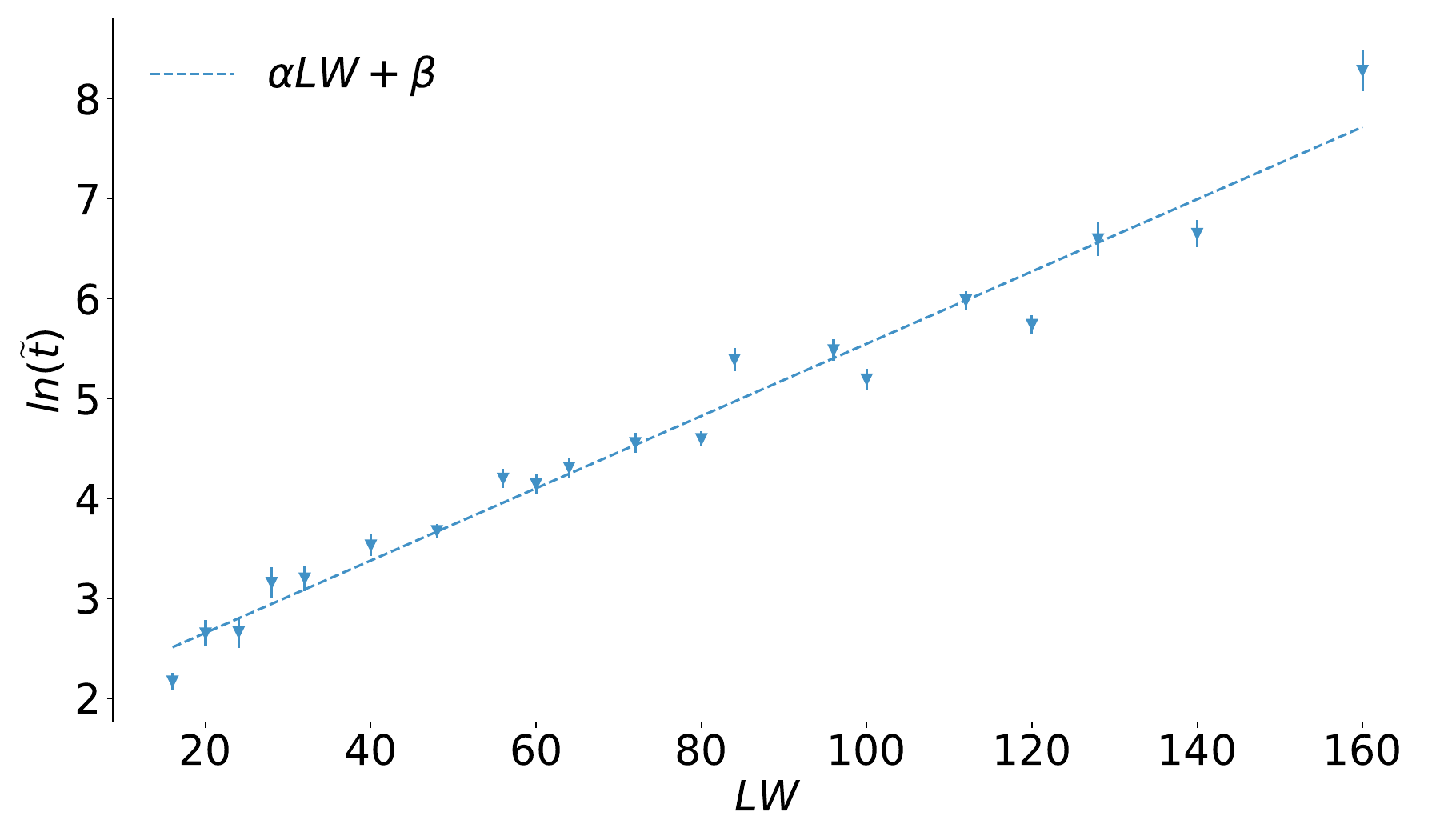}
\caption{$\ln(\widetilde{t})$ as a function of the product $LW$, and the linear fit providing the slope $\alpha$.}
\label{fig:slope}
\end{figure}

{To estimate the coefficient $\alpha$ in Eq.~\eqref{time:eqn}, we plot $ \ln \widetilde{t}$ as a function of the product $LW$, displaying all the points of Fig.~\ref{fig:relative} on one single curve shown in Fig.~\ref{fig:slope}, and apply a linear fit such that $\ln(\widetilde{t}) = \alpha LW + \beta$. From the fit we obtain $\alpha = 0.036 \pm 0.001$; Substituting this result in Eq.~\eqref{wutan:eqn} we find a threshold value $\tilde W \simeq 38.5$. This is larger than the estimate obtained below with the heat propagation (see Sec.~\ref{therma_front:sec}), but it is not a problem. Indeed, $\tilde t$ is only a lower bound to the slowest thermalization time. We know for sure that for $W>\tilde W$ the slowest thermalization time scales slowly enough that MBL is robust to avalanches; Below this threshold it could be robust or not. Nevertheless this analysis provides the important information that for $W>\tilde W$ MBL is a robust phase.} 

To gain a deeper understanding of the imbalance behavior, we focus now on the properties of the local magnetizations. 
\subsection{Thermalization front and local magnetizations}\label{therma_front:sec}
%
\subsubsection{Logarithmic propagation of the thermalization front}
\begin{figure}
    \includegraphics[width=\columnwidth]{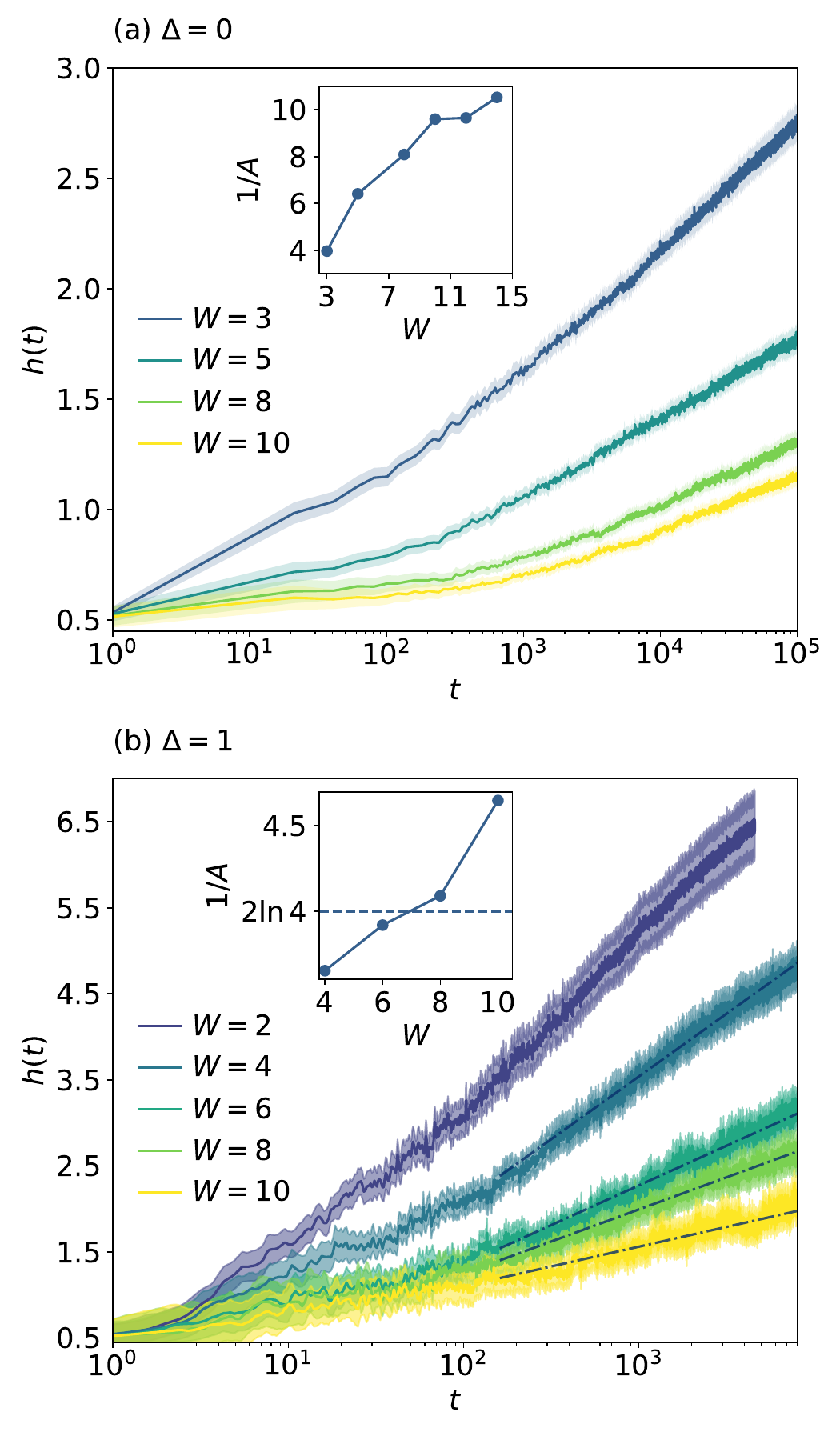}
  \caption{(Main panels) $h(t)$ versus $t$ in the Anderson ($\Delta=0$) case in panel (a) and interacting case ($\Delta = 1$) in panel (b). Notice the logarithmic scale on the horizontal axis. (Insets) $1/A$ versus $W$, where $A$ is the slope obtained linearly fitting $h(t)$ versus $\ln t$, applying the fit for times large enough that the linear in $\ln t$ regime has already set in. {The blue lines in the main panel (b) mark the linear fits}, and in the inset in panel (b) there is the reference value $1/A = 4\ln 2$, as defined in Eq.~\eqref{requiro:eqn}.  
  Numerical parameters in (a): $ L=100, N_r = 1200$. Numerical parameters in (b): $L = 16$, $N_r = 100$.}\label{spin_levo:fig}
\end{figure}

Let us consider the local magnetizations $\bruket{S_j^z}_{\gamma}(t) = \overline{\braket{\psi_t|\hat{S}_j^z|\psi_t}}$ for the evolution with a given $\gamma$. To see how thermalization propagates through the chain, let us set 
\begin{equation}\label{seggi:eqn}
  \delta S_j (t) = \left|\bruket{S_j^z}_{\gamma=0}(t)-\bruket{S_j^z}_{\gamma=1}(t)\right|\,,
\end{equation}
and define a thermalization length scale as
\begin{equation}\label{orcola:eqn}
  h(t) = \frac{\sum_{j=1}^{L}(j-1) \delta S_j(t)}{\sum_{j=1}^{L} \delta S_j (t)}\,,
\end{equation}
and study its behavior as a function of time. 
We plot $h(t)$ versus $t$ for the Anderson-model case in Fig.~\ref{spin_levo:fig}(a) and in the interacting case in Fig.~\ref{spin_levo:fig}(b). In the Anderson-model case, whatever the disorder strength $W$, we see that $h(t)$ logarithmically increases with time for $t$ large enough, as $ h(t) \sim A \ln(t)$. 
In the interacting case we see the same logarithmic increase for disorder strengths $ W > 2$. {We have checked that the result is converged in system size and does not depend on $L$ for the time interval we have numerically access to.} For $W=2$ we see a faster growth and a saturation due to finite-size effects.)
We obtain the slope $A$ with a linear fit of $h(t)$ versus $\ln t$, applying the fit for times such that the linear regime in $\ln t$ has already set in.
We plot the $1/A$ resulting from this fit as a function of $W$ in the insets of Fig.~\ref{spin_levo:fig}. In the Anderson-model case we see that $1/A$ irregularly increases with $W$, while in the interacting case the increase is more regular, slightly faster than linear.

\subsubsection{Relation with the behavior of the imbalance}
So we have found that the thermalization front propagates logarithmically. This gives rise to some interesting predictions on the behavior of $ \mathcal{I}_r(t)$. Because at time $t$ only a fraction $f(t)\sim A\ln(t)/L$ of the chain has thermalized, we can predict a behavior 
\begin{equation}\label{Irpre:eqn}
  \mathcal{I}_r(t)\sim A\ln(t)/L\,.
\end{equation}
{We have obtained this estimate as follows. Due to extensivity, let us write the long-time value of the noiseless imbalance $\overline{I}_{0}(t)=qL$ for some $q\in(0,1)$ and $t\gg 1/J$. If only a fraction of $f(t)$ sites has thermalized, we can roughly approximate $\overline{I}_{\gamma}(t)\sim L q[1-f(t)]$, and so we apply Eq.~\eqref{ira:eqn} and get Eq.~\eqref{Irpre:eqn}, valid for $t\gg 1/J$.}

\begin{figure}
\includegraphics[width = \columnwidth]{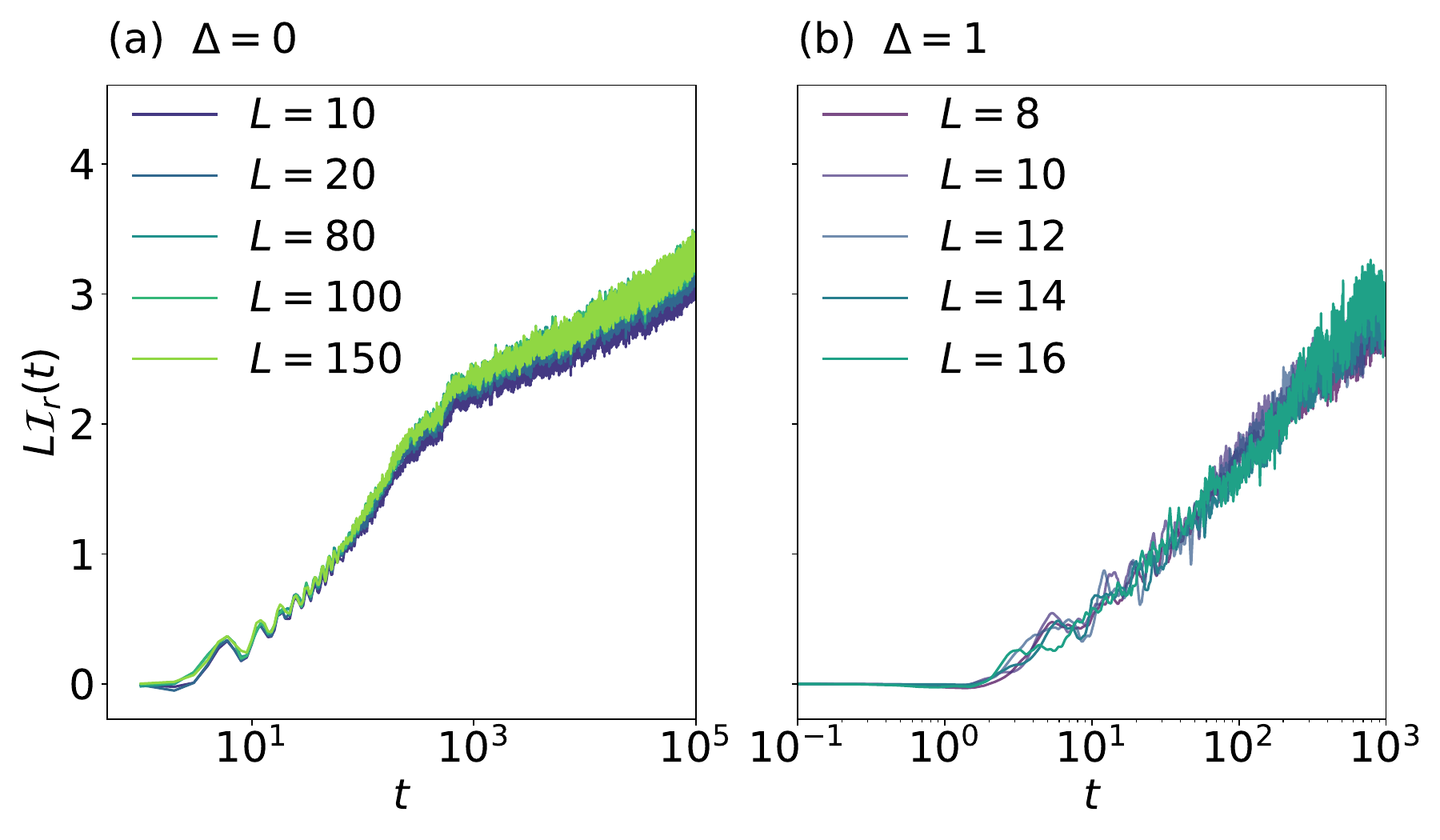}
  \caption{$L \mathcal{I}_r(t)$ versus $t$ for and $W=8$ in the Anderson ($\Delta=0$) case in panel (a) and in the interacting case in panel (b). Notice the logarithmic scale on the horizontal axis.}
  \label{rimba_levo:fig}
\end{figure}

{We have numerically checked} that this prediction is obeyed for large $L$ in the Anderson case as we show in Fig.~\ref{rimba_levo:fig}(a), and for smaller system sizes in the interacting case as we show in Fig.~\ref{rimba_levo:fig}(b). In both panels we fix $W$ and plot $L\mathcal{I}_r(t)$ versus $t$ for different values of $L$, with a logarithmic scale on the horizontal axis. We see that the rescaled curves tend to a limit curve, meaning that the scaling with $1/L$ holds for large system sizes. Furthermore, the limit curve increases linearly with time in logarithmic scale, consistently with Eq.~\eqref{Irpre:eqn}. This finding implies an exponential scaling with the system size of the thermalization time of the imbalance, as the one shown in Fig.~\ref{fig:relative}(b).

As we can see in the inset of Fig.~\ref{spin_levo:fig}(b), the behavior of $1/A$ is consistent with a linear increase in $W$, so we can write $1/A \sim \beta W$ for some $\beta > 0$. Substituting $1/A = \beta W$ in Eq.~\eqref{Irpre:eqn} and imposing $\mathcal{I}_r (\widetilde{t}) = r_{\rm th}(=0.17)$ 
we get
\begin{equation}
  \widetilde{t}\sim\exp\left(r_{\rm th}\beta L W\right)\,,
\end{equation}
consistent with Eq.~\eqref{time:eqn}.
\subsubsection{Slowest thermalization time and robustness to avalanches}
Let us see what the results above imply for robustness of localization against avalanches. In Sec.~\ref{sec:imbalance} we have derived the threshold value in Eq.~\eqref{wutan:eqn}, which implies the robustness of MBL for $W$ large enough. Let us now assume to average over infinite disorder/randomness realization and observe that Eq.~\eqref{seggi:eqn} implies
\begin{equation}
  \delta S_j(\infty)=\lim_{t\to\infty}\delta S_j (t) = \lim_{t\to\infty}\left|\bruket{S_j^z}_{\gamma=0}(t)\right|\,,
\end{equation}
because $\lim_{t\to\infty}\bruket{S_j^z}_{\gamma=1}(t)=0$ $\forall j=1,\,\ldots,\,L$ due to thermalization. In periodic boundary conditions (PBC), due to the symmetries of the problem averaged over disorder, both $\left|\bruket{S_j^z}_{\gamma=0}(t)\right|$ and $\delta S_j(\infty)$ are independent of $j$, (let us write $\delta S_j(\infty)\equiv\delta S $), so that 
\begin{equation}
  \lim_{t\to\infty}h(t) = \frac{\delta S\sum_{j=1}^L(j-1)}{\delta S\sum_{j=1}^L 1} = \frac{L-1}{2}\quad\text{ for PBC}\,.
\end{equation}

The open boundary conditions affect the behavior of the $\left|\bruket{S_j^z}_{\gamma=0}(t)\right|$, and then of the $\delta S_j(\infty)$, for a finite length $l_{b}$ near the boundaries. This length $l_{b}$ is finite and does not scale with $L$ because we assume the system with $\gamma=0$ to be localized (Anderson or MBL). So we can write in our case
\begin{equation}
  \lim_{t\to\infty}h(t) = \frac{L-1}{2}+\mathcal{O}(l_{b}) \quad \text{ for OBC}\,.
\end{equation}

We can say that all local magnetizations have thermalized, at time $t^*$, when $h(t^*) = \lim_{t\to\infty}h(t)$. Using that for $t \gg 1/J$ the length scale $h(t)\sim A\ln(t)$, we get with some simple algebra the thermalization time $t^*$ as
 \begin{equation}\label{turga:eqn}
   t^*\sim\exp\left(\frac{L}{2A}\right)\,,
 \end{equation}
with some multiplicative constants in front that are irrelevant for the scaling with $L$. In order to have robustness against avalanches, we must have a scaling faster than $4^L$. We meet this condition if $\exp\left(\frac{1}{2A}\right)>4$ or
\begin{equation}\label{requiro:eqn}
 \frac{1}{A} > 2 \ln 4\,.
\end{equation}

In the inset of Fig.~\ref{spin_levo:fig}(a), we observe that the Anderson model consistently satisfies this requirement within the parameter range we investigate. The Anderson model is inherently integrable and unable to generate ergodic inclusions. However, if an ergodic inclusion is introduced into the model, it remains resistant to thermalization induced by avalanches within this parameter range.

In contrast, the MBL model satisfies Eq.~\eqref{requiro:eqn} provided that $W$ exceeds a threshold value $W^* \gtrsim 7$, as shown in the inset of Fig.~\ref{spin_levo:fig}(b). For $W > W^*$, MBL remains robust. For $W < W^*$, we cannot make definitive claims about stability because $t^*$ in Eq.~\eqref{turga:eqn} only offers a lower bound on the slowest thermalization time $t_s$.

\section{Conclusion and perspectives}\label{conc:sec}
In conclusion we have studied a many-body or Anderson localized model coupled to a $T=\infty$ thermal bath by one end. The coupling is described by a Lindbladian, and to numerically study it we have used a quantum-trajectory approach. More specifically, we have used a unitary unraveling such that the dynamics is an average over many realizations of a unitary Schr\"odinger evolution with noise. In the case of Anderson localization this approach allows to reach quite large system sizes, due to the Gaussian form of the state along each quantum trajectory.

We have first numerically studied the dynamics of the imbalance, a global quantity widely used to assess the presence of localization. We have defined its thermalization time as the time beyond which the normalized difference of the imbalance with and without thermal bath goes beyond a given threshold, and we see that in the MBL case this thermalization time exponentially increases with the strength of the disorder and the size of the system. We have shown that this result implies that the MBL is robust against the avalanche instability when the disorder strength goes beyond a given threshold. 

We have then focused on the heat propagation through the system considering how a thermalization front propagated. We have estimated the extension of the already thermalized part of the chain, defining a thermalization length scale using the onsite magnetizations, and found that it logarithmically increases with time, {in agreement with the existing analytical estimates~\cite{toniolo2024}}. This is true both for the Anderson and the MBL case and we could use it to lower bound the slowest thermalization time with a quantity exponentially scaling with the system size. We have found that in the strong-disorder regime this scaling is fast enough that the system is robust to avalanches. 

As a perspective for future work, we can focus our attention on the density-density correlator~\cite{PhysRevLett.114.100601}, a quantity whose logarithmically slow propagation properties are known in the noisy Anderson case~\cite{Talia_Bar_Lev_22,devendra_24}, and see how this behavior is changed in the MBL case. Second, we can perform a similar analysis on the OTOC (that for MBL systems has been considered in~\cite{PhysRevB.95.165136}) to see how the thermalization front affects the scrambling properties of the system. Finally, we can study more deeply the relation between the strength of localization of the integrals of motion, the logarithmic thermalization front, the scaling of the slowest thermalization time, and the robustness of MBL to avalanches, {also in MBL models with long-range interactions~\cite{PhysRevB.93.245427,PhysRevLett.113.243002,PhysRevLett.128.146601}, where the behavior of the thermalization length scale has not yet been explored, to the best of our knowledge.}

\begin{acknowledgments}
 We acknowledge useful discussions with D.~Singh Bhakuni and useful comments on the manuscript by D.~Huse, N.~Laflorencie, D.~Luitz and D.~Toniolo. M.\,C., G.\,P., P.\,L. and A.\,R. acknowledge financial support from PNRR MUR Project PE0000023-
NQSTI and 
computational resources from MUR, PON “Ricerca e Innovazione 2014-2020”, under Grant No. PIR01\_00011
- (I.Bi.S.Co.).
G.\,P.\ acknowledges computational resources
from the CINECA award under the ISCRA initiative. This work was furthermore supported by the European
Union’s Horizon 2020 research and innovation programme under Grant Agreement No 101017733, by
the MUR project CN\_00000013-ICSC (P.\,L.), and by
the QuantERA II Programme STAQS project that
has received funding from the European Union’s Horizon 2020 research and innovation program.
\end{acknowledgments}

\appendix

\section{Derivation of the Lindblad equation form the quantum-trajectory scheme}\label{derivation:sec}
Let us start with the Schr\"odinger equation with the Hamiltonian Eq.~\eqref{hmbl:eqn}
\begin{equation}
  i\frac{\ud}{\ud t}\ket{\psi(t)} = [\hat{H}_0 + \gamma\xi(t)\hat{S}_1^z]\ket{\psi(t)}\,.
\end{equation}
We can write it as
\begin{align}\label{oknus:eqn}
 &\ket{\psi(t+\Delta t)}\bra{\psi(t+\Delta t)}  =\ket{\psi(t)}\bra{\psi(t)} \\&\quad{}-i\Delta t[\hat{H}_0,\ket{\psi(t)}\bra{\psi(t)}] 
   -i \Delta W_t \gamma [\hat{S}_1^z,\ket{\psi(t)}\bra{\psi(t)}]\,, \notag
\end{align}
where $\braket{\Delta W_t} = \Delta t$ and $\braket{\Delta W_t\Delta W_{t'}} = \delta_{t,\,t'}\Delta t$ are Gaussian uncorrelated variables. Let us also write
\begin{align}
  \ket{\psi(t)}\bra{\psi(t)} &= \ket{\psi(t-\Delta t)}\bra{\psi(t-\Delta t)}\notag\\
   &-i\Delta t[\hat{H}_0,\ket{\psi(t-\Delta t)}\bra{\psi(t-\Delta t)}] \\
   &-i \Delta W_t \gamma [\hat{\sigma}_1^z,\ket{\psi(t-\Delta t)}\bra{\psi(t-\Delta t)}]\,,\notag
\end{align}
and substitute it in the last term of Eq.~\eqref{oknus:eqn}. We get many terms. Averaging over randomness and defining $\hat{\rho}_t=\big\langle\varrho(t)\big\rangle$ with $\varrho(t)\equiv\ket{\psi(t)}\bra{\psi(t)}$, we get
\begin{align}
  \hat{\rho}_{t+\Delta t} &= \hat{\rho}_t - i \Delta t [\hat{H}_0,\hat{\rho}_t]  \notag \\ &\quad {}-(\overline{\Delta W_t\Delta W_{t}})\gamma^2 [\hat{S}_1^z,[\hat{S}_1^z,\hat{\rho}_{t-\Delta t}]]
   + o(\Delta t)\,.
\end{align}
Using $\braket{\Delta W_t\Delta W_{t}} = \Delta t$ and going in the limit $\Delta t \to 0$ we get
\begin{equation}
  \frac{\ud}{\ud t}\hat{\rho}_t =- i [\hat{H}_0,\hat{\rho}_t] - \gamma^2 [\hat{S}_1^z,[\hat{S}_1^z,\hat{\rho}_{t}]]\,.
\end{equation}

\section{Case of the Anderson model}\label{anderson:sec}
Applying the Jordan-Wigner transformation, the model in Eq.~\eqref{eq:xxzmodel} becomes the spinless-fermion model
\begin{equation}
  \hat{H}_0 = \sum_{j=1}^L h_j \hat{n}_j +  J \sum_{j=1}^{L-1} \left[\frac{1}{2}( \hat{c}_j^\dagger\hat{c}_{j+1} + {\rm H.\,c.}) + \Delta \hat{n}_j \hat{n}_{j+1} \right]
  \label{eq:Ham0}
\end{equation}
where $\hat c^{(\dagger)}_j$ are anticommuting fermionic operators acting
on the $j$-th site. The number of fermions $N\equiv\sum_{j=1}^L\hat{n}_j$ is conserved. In the Anderson-model case we have $\Delta = 0$ and we get a quadratic fermionic Hamiltonian, so that the time evolved state $\ket{\psi(t)}$ along each trajectory can be cast in the form of a generic Gaussian state (Slater determinant).
The full information of such state is contained in a $L \times N$ matrix ${\bf W}(t)$, defined by
\begin{equation}
  \ket{\psi(t)} = \prod_{k=1}^N \left[ \, \sum_{j=1}^L \big[ W(t) \big]_{j,k} \, \hat c_j^\dagger\right] \ket{\Omega}\,,
  \label{eq:Psi_t}
\end{equation}
{where $\ket{\Omega}$ is the vacuum of the fermionic operators $\hat{c}_j$.} Because the number of fermions is conserved and we initialize with the N\'eel state we have $N=L/2$. At initial time, the matrix $L \times N$ matrix defining the N\'eel state is given by $[W(0)]_{k,l} = \sum_{j=1}^{L/2} \delta_{2j,l}\delta_{j,k}$. The discrete evolution step (see Eq.~\eqref{unno:eqn})
\begin{equation}
  \ket{\psi({t+\tau})} =  \nep^{-i\eta_n\gamma\hat{S}_1^z}\nep^{-i\tau\hat{H}_0}\ket{\psi(t)}
\end{equation}
is simply translated in the dynamics of the matrix ${\bf W}(t)$ as
\begin{equation}
  {\bf W}(t+\tau) = {\bf K}_n\nep^{-i\tau {\bf Q}} {\bf W}(t)\,,
\end{equation}
{where the noisy step is implemented through the $L\times L$ diagonal matrix ${\bf K}_n$ with matrix element $[K_n]_{i,j}=\delta_{i,j} + \delta_{1,i}\delta_{1,j}(\nep^{-i\eta_n\gamma}-1)$, while the $L\times L$ matrix ${\bf Q}$ implementing the action of the Anderson Hamiltonian has matrix elements $[Q]_{i,j} = \delta_{i,j} h_j + \frac{J}{2}(\delta_{i,j+1} + \delta_{i,j-1})$.} Thanks to the validity of Wick's theorem, one can use the matrix ${\bf W}(t)$ to obtain the expectation of any observable, and even the entanglement entropy, as clarified in~\cite{DeLuca2019}. In this way one can numerically reach quite large system sizes, up to $L\sim\mathcal{O}(10^2)$. 
%
\section{{Particle-number non-conserving case}}\label{relax:app}
\begin{figure}
    \centering
    \includegraphics[width=\columnwidth]{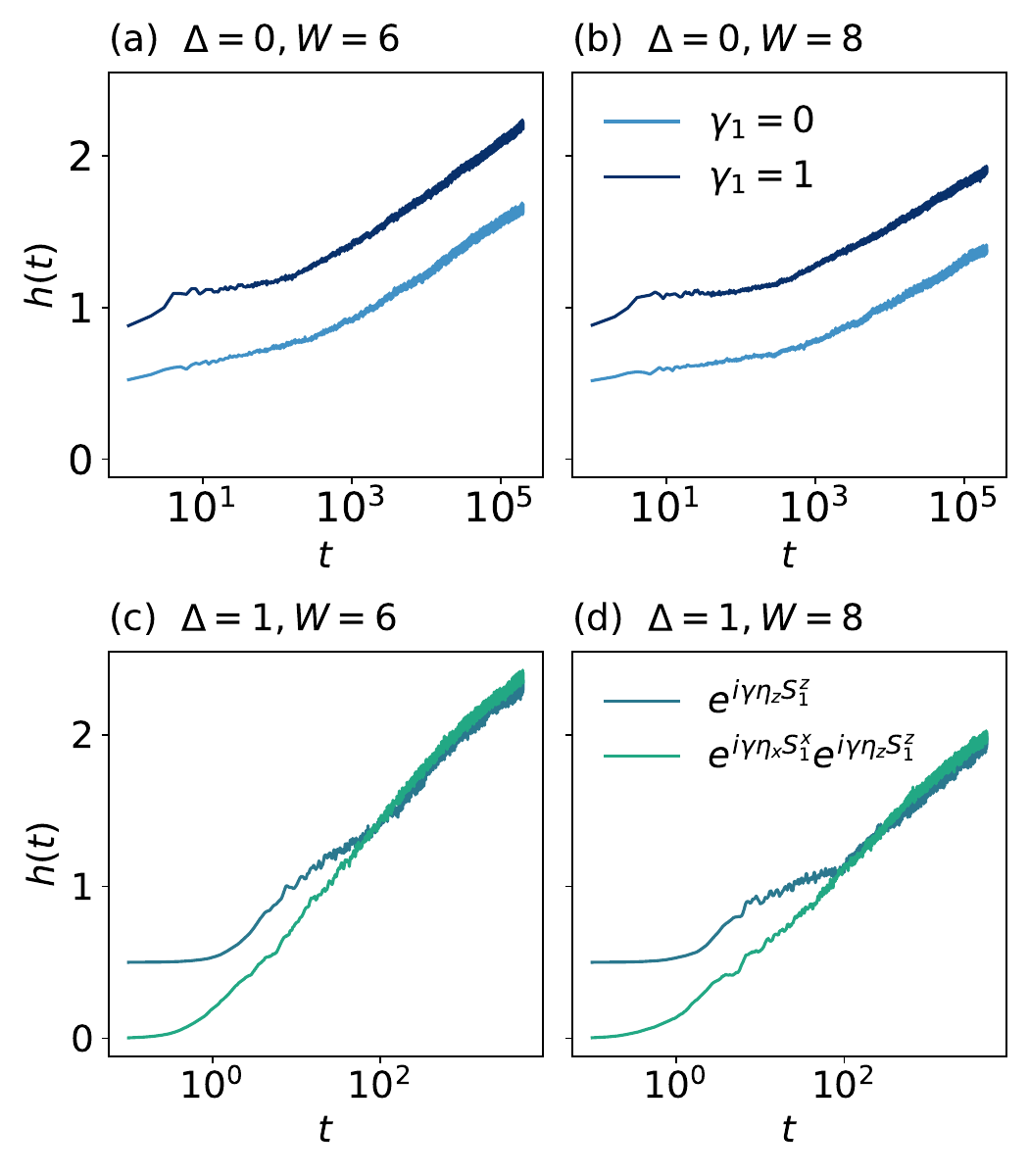}
    \caption{(Panels a, b) Examples of comparison of $h(t)$ versus $t$ with the particle-number conserving ($\gamma_1 = 0$) and nonconserving ($\gamma_1=1$) noises for different values of $W$ in the Anderson-model case ($\Delta = 0$). Other numerical parameters: $\gamma=1,\,L=10,\,\,N_{r} = 1200$. (Panel c, d) Examples of comparison of $h(t)$ versus $t$ with the particle-number conserving noise Eq.~\eqref{unno:eqn} and nonconserving noise Eq.~\eqref{unno1:eqn} for different values of $W$ and $\Delta = 1$. Other numerical parameters: $\gamma=1,\,L = 8,\,N_r = 576$.}
    \label{fig:ht5678}
\end{figure}

{The coupling to the bath of Eq.~\eqref{lindblad:eqn} conserves the total spin, which is the particle number in the fermionic representation of Eq.~\eqref{eq:Ham0}. Breaking this conservation does not change very much, due to the propagation of heat being constrained by a logarithmic light cone~\cite{toniolo2024}.
In order to check this, let us focus on the Anderson model ($\Delta =0$), and apply to it a different particle-number nonconserving boundary noise
\begin{equation}\label{hambl:eqn}
  \hat{H}_{\gamma}(t) = \hat{H}_0 + \gamma\xi_1(t)\hat{S}_1^z + \gamma_1\xi_2(t)\left(\hat{S}_1^+\hat{S}_2^++\hat{S}_1^-\hat{S}_2^-\right)\,,
\end{equation}
where $\xi_1(t)$ and $\xi_2(t)$ are uncorrelated Gaussian random processes $\braket{\xi_j(t)\xi_l(t')} = \delta_{j\,l}\delta(t-t')$. Averaging over the noise, one gets the Lindblad equation
\begin{align} 
    \dot{\hat{\rho}}(t) &= - i [ \hat{H}_0, \hat{\rho}(t) ] - \gamma^2 [\hat{S}^z_1 ,[\hat{S}^z_1, \hat{\rho}(t)]]\nonumber\\
     &-\gamma_1^2\left[\left(\hat{S}_1^+\hat{S}_2^++\hat{S}_1^-\hat{S}_2^-\right) ,\left[\left(\hat{S}_1^+\hat{S}_2^++\hat{S}_1^-\hat{S}_2^-\right), \hat{\rho}(t)\right]\right]\,,
\end{align}
The dynamics in Eq.~\eqref{hambl:eqn} can be numerically studied by applying the Jordan-Wigner transformation and using the Bogoliubov-de Gennes formalism~\cite{santoro_scipost}. If we discretize the time with a step $\tau$, Trotterize, and apply the Jordan-Wigner transformation we get the discrete time-step evolution operator
\begin{equation} 
  \hat{U}_n = \nep^{-i\gamma\eta_{n}^{(1)}\hat{n}_1^z}\nep^{-i\gamma_1\eta_{n}^{(2)}\left(\opcdag{1}\opcdag{2}-\opc{1}\opc{2}\right)}\nep^{-i\tau\hat{H}_0}\,,
\end{equation}
with
\begin{equation}
  \hat{H}_0 = \sum_{j=1}^L h_j \hat{n}_j +  \frac{J}{2} \sum_{j=1}^{L-1} ( \hat{c}_j^\dagger\hat{c}_{j+1} + {\rm H.\,c.}) \,,
\end{equation}
and $\eta_{n}^{(1)}$, $\eta_{n}^{(2)}$ are uncorrelated zero-mean Gaussian random variables such that $\braket{\eta_{n}^{(j)}\eta_{n'}^{(k)}}=\delta_{j\,k}\delta_{n\,n'}\tau$. Using the Bogoliubov formalism we introduce the fermionic operators $\hat{\gamma}_\alpha(t)=\sum_{j=1}^L\left(U^*_{j\,\alpha}(t)\opc{j}+V^*_{j\,\alpha}(t)\opcdag{j}\right)$ such that $\hat{\gamma}_\alpha(t)\ket{\psi(t)}=0$, and the coefficients $U_{j\,\alpha}(t)$, $V_{j\,\alpha}(t)$ form a $2L\times L$ matrix that obeys the relation
\begin{align}
  \left(\begin{array}{c} {\bf U}(t+\tau)\\{\bf V}(t+\tau)\end{array}\right)=&
  \exp\left[-i\gamma\eta_{n}^{(1)}\left(\begin{array}{cc} {\bf A}&\boldsymbol{0}\\\boldsymbol{0}&-{\bf A}\end{array}\right)\right]\nonumber\\
  &\exp\left[-i\gamma_1\eta_{n}^{(2)}\left(\begin{array}{cc} \boldsymbol{0}&{\bf B}\\-{\bf B}&\boldsymbol{0}\end{array}\right)\right]\nonumber\\
  &\exp\left[-i\tau\left(\begin{array}{cc} {\bf Q}&\boldsymbol{0}\\\boldsymbol{0}&-{\bf Q}\end{array}\right)\right]\left(\begin{array}{c} {\bf U}(t)\\{\bf V}(t)\end{array}\right)\,,
\end{align}
where the $L\times L$ matrices ${\bf A}$, ${\bf B}$, and ${\bf Q}$ have matrix elements
\begin{align}
  [A]_{jl} &= \delta_{j1}\delta_{l1}\nonumber\\
  [B]_{jl} &= \delta_{j1}\delta_{l2}\nonumber\\
  [Q]_{jl}    &= h_j\delta_{jl} + \frac{J}{2}(\delta_{j\,l+1}+\delta_{j\,l-1})\,.
\end{align}
The on-site magnetizations at time $t$ are given by
\begin{equation}
  \braket{\hat{S}_j^z}_t = \sum_{\alpha}|V_{j\alpha}(t)|^2-1/2\,.
\end{equation}
We compare for some values of $W$ the evolution of $h(t)$ under particle-number conserving noise ($\gamma_1=0$) and the one under particle-number nonconserving noise ($\gamma_1=1$). Putting the logarithmic scale on the horizontal axis, we see that in each cases the two curves tend to straight lines with the same slope [see Fig.~\ref{fig:ht5678}(a,b)]. Therefore, the value of the slope $A$ (needed for estimating $t^*$) is the same with and without particle-number conservation. The physical reason is that the propagation is bounded by the same logarithmic light cone with and without particle conservation~\cite{toniolo2024}.}
%

{We can do a similar analysis for the interacting case with $\Delta = 1$. In this case we apply the following particle-number nonconserving boundary noise
\begin{equation}
   \hat{H}_{\gamma}(t) = \hat{H}_0 + \gamma\xi_z(t)\hat{S}_1^z + \gamma\xi_x(t)\hat{S}_1^x\,.
\end{equation}
So we must compare the dynamics induced by the operator Eq.~\eqref{unno:eqn} with the one induced by
\begin{equation}\label{unno1:eqn}
 \hat{U}_n' = \nep^{-i\eta_n^{x}\gamma\hat{S}_1^x}\nep^{-i\eta_n^{z}\gamma\hat{S}_1^z}\nep^{-i\tau\hat{H}_0}\,,
\end{equation}
where $\eta_n^{z}$ and $\eta_n^{x}$ are Gaussian zero-average random variables such that $\braket{\eta_n^{\alpha}\eta_m^{\beta}}=\tau\delta^{\alpha\beta}\delta_{mn}$. We show some examples of comparison of $h(t)$ obtained with the two types of noise, for corresponding parameters of $\hat{H}_0$, in Fig.~\ref{fig:ht5678}(c,d). We see that after a transient the two types of noise provide the same curve.}
\section{Scaling of the slowest thermalization time and robustness of localization against avalanches}\label{ava_come_lava:sec}
\begin{figure}
  \begin{center}
  \includegraphics[width=\columnwidth]{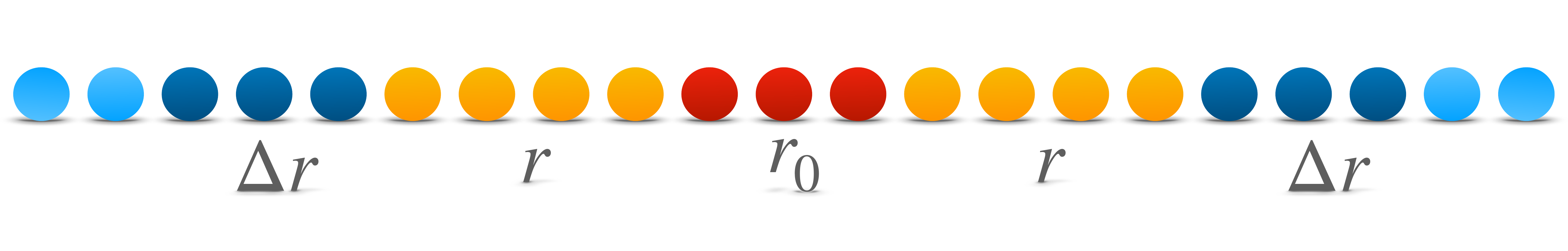}
  \caption{Scheme of an MBL chain for the analysis of robustness of MBL to avalanches. The red region is the ergodic inclusion, the yellow region the thermalized part of the chain, the blue region the part of the chain not yet thermalized. We focus on a subset of the latter (dark blue) that is in contact with the thermalized region and is made by two sectors, each of length $\Delta r$.}\label{scheme:fig}
  \end{center}
\end{figure}

In this section we explain why, in order for localization to be robust against avalanches, the slowest thermalization time should scale faster than $4^L$ when one end of the system is coupled to a thermal bath. The thing has already been clearly explained in~\cite{PhysRevB.105.174205,PhysRevB.106.L020202}, and we add this section just for completeness. Let us consider Fig.~\ref{scheme:fig} depicting an MBL chain. We have an ergodic inclusion of size $r_0$ (red region), and a part that has already thermalize (yellow region) made by two sectors, with the same size $r$ (we assume the avalanches to act symmetrically). The rest of the system is already localized (blue region) and we want to see if thermalization can further propagate there by avalanches.
With this aim, we focus on a part of the localized region in contact with the ergodic region, made by two sectors of length $\Delta r$ (dark blue region). We ask ourselves if this region thermalizes due to the contact with the yellow region, that's to say if the avalanche propagates. This can happen if each sector in the dark-blue region (let's say the right one -- they are equal) thermalizes fast enough. So the slowest thermalization time $t_s$ of each dark-blue sector must be much shorter than the inverse gap $1/\delta$ of the region obtained joining the red, the yellow and the dark blue regions. The point is that thermalization in each dark-blue sector must be faster than the time needed to the part of the chain involved in the thermalizing dynamics (red+yellow+dark-blue) to express finite-size revivals and other quantum dynamical effects connected with the discreteness of the spectrum that hinder thermalization.
On general grounds~\cite{PhysRevB.105.174205,PhysRevB.106.L020202} (and we numerically verify it also in our work) one has $t_s\sim \kappa^{\Delta r}$ for some $\kappa>0$. Being this a spin 1/2 system one has $\delta=2^{-(r_0+2r+2\Delta r)}$. Imposing $t_s \ll 1/\delta$, and asking that this condition should be verified also in the limit of $\Delta r \gg r$, one finds
\begin{equation}
  \kappa < 4\,.
\end{equation}
By contrast, if $\kappa > 4$ MBL is robust to avalanches. So it is very important to know $\kappa$. In order to find it, we have done as biologists do when move a portion of neural tissue from {\em in vivo} to {\em in vitro}~\cite{PhysRevB.105.174205,PhysRevB.106.L020202}: We have cut away the right blue region from the system in Fig.~\ref{scheme:fig}, we have connected its left site to a thermal bath simulating the yellow thermalized region in Fig.~\ref{scheme:fig}, and we have studied the scaling of the slowest thermalization time with $\Delta r$. We have renamed $\Delta r$ as $L$ for merely aesthetic reasons.


\begin{thebibliography}{94}
\bibitem [{not({\natexlab{b}})}]{nothreshold:note}%
  \BibitemOpen
  \href@noop {} {} \bibinfo {note} {For the smaller values
  of $L$, where relaxation times are shorter, we can use larger threshold, and
  get clearer exponential increases with $W$ than the ones shown in
  Fig.~\ref{fig:relative}(a).}\BibitemShut {Stop}%
\end{thebibliography}
%
%
\end{document}